\documentclass[traditabstract
,letter
]{aa}  
\usepackage{graphicx,amsmath}
\usepackage{color}
\usepackage{txfonts}

\begin{document}
\title{Exocomets in the circumstellar gas disk of HD\,172555}

\author{
F.~Kiefer\inst{1,2}
\and
A.~Lecavelier des Etangs\inst{1,2}
\and
J.-C~Augereau\inst{3}
\and
A.~Vidal-Madjar\inst{1,2}       
\and 
A.-M.~Lagrange\inst{3}
\and
H.~Beust\inst{3}
 }
   


   \institute{
   CNRS, UMR 7095, 
   Institut d'astrophysique de Paris, 
   98$^{\rm bis}$ boulevard Arago, F-75014 Paris, France
   \and
   UPMC Univ. Paris 6, UMR 7095,
   Institut d'Astrophysique de Paris, 
   98$^{\rm bis}$ boulevard Arago, F-75014 Paris, France
   \and
   UJF-Grenoble 1/CNRS-INSU, 
   Institut de Plan\'etologie et dAstrophysique (IPAG) UMR 5274, 38041 Grenoble, France
    }
   
   \date{} 
 
  \abstract{

The source HD172555 is a young A7V star surrounded by a debris disk with a gaseous component. 
Here, we present the detection of variable absorption features detected simultaneously 
in the Ca\,II K and H doublet lines (at $\lambda$3,933\AA\ and $\lambda$3,968\AA ). 
We identified the presence of these absorption signatures at four different epochs in the 129~HARPS high-resolution spectra gathered between 2004 and 2011. 
These transient absorption features are most likely due to Falling Evaporating Bodies 
(FEBs, or exocomets) that produce absorbing gas observed transiting in front of the central star. 
We also detect a stable Ca\,II absorption component at the star's radial velocity. With no corresponding 
detection in the Na\,I line, 
the resulting very low upper limit for the Na\,I/Ca\,II ratio suggests that this absorption is due to circumstellar gas. 

%
}


   \maketitle
%


\section{Introduction}
 \label{Introduction}

The source HD172555 is an A7V-type star in the $\beta$\,Pictoris moving group and harbors a dusty and gaseous circumstellar (CS) disk (Cote 1987; Lisse et al.\ 2009; Riviere-Marichalar et al.\ 2012). Lisse et al.\ (2009) and Johnson et al.\ (2012) have proposed that the dust was recently produced by a catastrophic collision between planetary mass bodies, which would have happened at around 5 au from the star. 
This scenario is consistent with the mid-infrared interferometric observations of the spatial distribution of dust detected at distances larger than 1\,au from the central star (Smith et al.\ 2012). Up to now, the search for massive planets in the system have yielded only negative results (Quanz et al.\ 2011). 
The most striking recent discovery in this young circumstellar disk is the detection of [OI] emission at 63.2\,$\mu$m (Riviere-Marichalar et al.\ 2012), which shows that most of the mass of the disk must be in the gaseous phase. These works suggest that the combination of a dust debris disk with a gaseous circumstellar disk surrounding an A-type star a dozen million years old makes the HD172555 system much resembling $\beta$\,Pictoris.

The gaseous component of the $\beta$\,Pic disk was detected in Ca\,II (Hobbs et al.\ 1985); it presents an anomalously low  Na\,I/Ca\,II ratio with a column density ratio $N(\text{Na\,I})/N(\text{Ca\,II})$$\sim$0.03 and an equivalent width ratio $EqW(\text{Na\,I}\,D_2)/EqW(\text{Ca\,II}\, K)$$\sim$0.1  (Vidal-Madjar et al.\ 1986). 
Furthermore, surveys of the $\beta$\,Pic spectrum revealed the presence of variable and random additional absorptions, most often redshifted in the Ca\,II lines (Ferlet et al.\ 1987). These features are well interpreted in terms of Falling Evaporating Bodies (herafter FEBs, see review in Vidal-Madjar et al.\ 1998) or, in other words, exocomets. 
 
With the $\beta$\,Pic analogy in mind, we searched for CS gas and for any possible variable spectral signatures in the 129 spectra of HD172555 collected from 2004 to 2011 using the \emph{HARPS} spectrograph. 
We present the analysis of this data set in Sect.~\ref{Data analysis}. 
A stable absorption component is detected in the Ca\,II doublet at the radial velocity 
of the star and is most likely of circumstellar origin (Sect.~\ref{The stable gaseous component}). We also 
present the detection of sporadic absorptions with typical characteristics of FEBs (Sect.~\ref{FEB}). The results are discussed in Sect.~\ref{Discussion}.

\section{Data analysis}
\label{Data analysis}

The spectra were obtained from 2004 to 2011 
with the \emph{HARPS} spectrograph (R$\sim$115,000) installed at the La Silla 3.6m telescope (Table~\ref{tab:time_tab}). 
We focused our attention on the Ca\,II stellar lines which are the most sensitive to the transit of gaseous clouds like the ones produced by exocomets in $\beta$ Pic. To ensure that any detected variation is of astrophysical origin and
not due to variations in the wavelength or in the flux calibrations over the years, we compared the spectra 
in the region of the very steep Na\,I interstellar lines; this check confirms the tremendous 
stability of the instrument during the observation campaign (as expected for this spectrograph
aimed at detecting minute variations in the star's radial velocities), and thus provides very high confidence 
to the reality of the detected spectral evolutions (see below). 

From the 129~\emph{HARPS} spectra, we calculated the reference spectrum ($F_{ref}$) of HD172555. 
It is used to search for variable absorption features and is obtained as follows. 
First, we assume that in the absence of variable absorption features the noise $\Delta F$ 
in the flux measurement is Gaussian, fluctuating around the real reference spectrum. 
We checked that its $RMS$ is proportional to $\sqrt{F}$ with a wavelength independent factor, for which we obtained a reliable estimation in FEB-free regions.  Second, at each wavelength bin, the number N of flux measurements is equal to the total number of spectra. 
Since we assumed that $\Delta F$ follows a Gaussian distribution, the $k^{\rm th}$ highest flux value $f_k$ is an estimator of the $\alpha_k\, RMS$ level at which probability $P(\Delta F <\alpha_k\, RMS)$ is equal to $1-k/N$. 
By simple subtraction of $f_k$ from the $\alpha_k \, RMS$ value we thus obtain an estimation $F_{ref, k}$ of the reference spectrum. The final reference spectrum is the mean of all these estimations obtained by varying $k$ from 3 to 10. 

\begin{figure}[htb!]
\hbox{\centering
\includegraphics[angle=90,width=0.5\columnwidth]{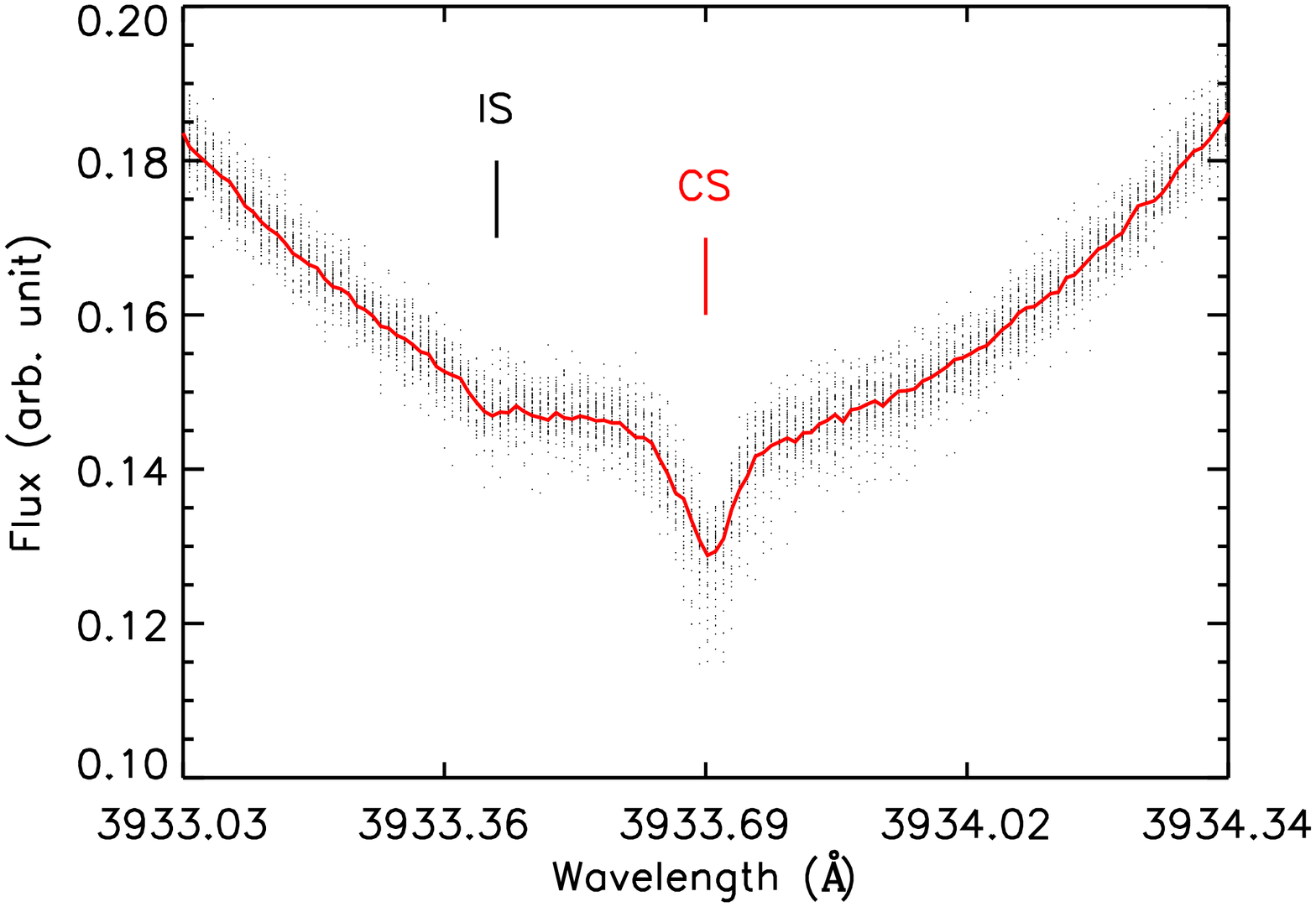}
\includegraphics[angle=90,width=0.5\columnwidth]{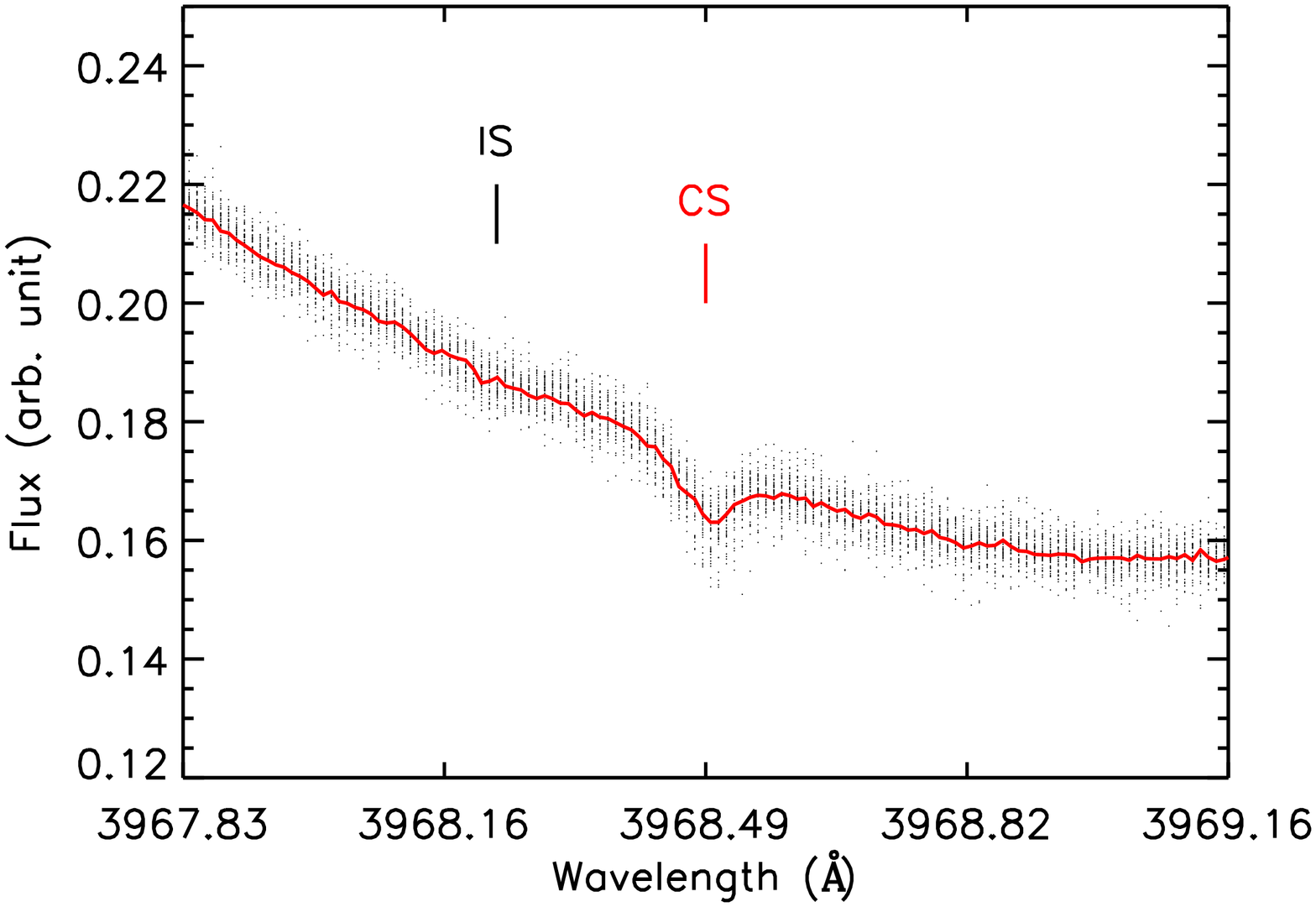}
}\caption
[Ca II reference spectrum]{
\label{fig:ref_spec} 
Plot of the Ca II reference spectrum of HD172555 (red line) in the Ca\,II K and H lines 
(left and right panels, respectively). 
All of the 129~spectra are represented 
by tiny black dots. The position of the interstellar (IS) and circumstellar (CS) absorptions  
are identified by vertical ticks. Wavelengths are expressed in the heliocentric rest frame. 
}
\end{figure}

\begin{table*}[htb]
\begin{center}
\begin{tabular}{ccccccc}
\hline
Line  & Rad. Vel.     & \multicolumn{4}{c}{Equivalent Width} & Na\,I/Ca\,II Ratio \\
Origin& (km\,s$^{-1}$) & \multicolumn{4}{c}{(m\AA)} & (Log$_{10}$(D2/K)\ ) \\
      & & Ca\,II K & Ca\,II H & Na\,I D2 & Na\,I D1 &  \\
\hline
IS & -19 & 0.87 $\pm$ 0.08 & 0.33 $\pm$ 0.05 &  1.02 $\pm$ 0.05 & 0.58 $\pm$ 0.04 & 0.07 $\pm$ 0.10 \\
CS & 2 & 8.33 $\pm$ 0.17 & 4.66 $\pm$ 0.12 & $<$0.31 (5-$\sigma$) & $<$0.31 (5-$\sigma$) & $<$-1.4 \\
\hline
\end{tabular}
 \caption{
 \label{tab:EqW}
 Properties of the absorption features detected in the Ca\,II and Na\,I doublets. 
 Radial velocities are given in heliocentric rest frame.
 }
\end{center}
\end{table*}

\begin{figure}[tbh!]
\hbox{
\includegraphics[angle=90,width=0.48\columnwidth]{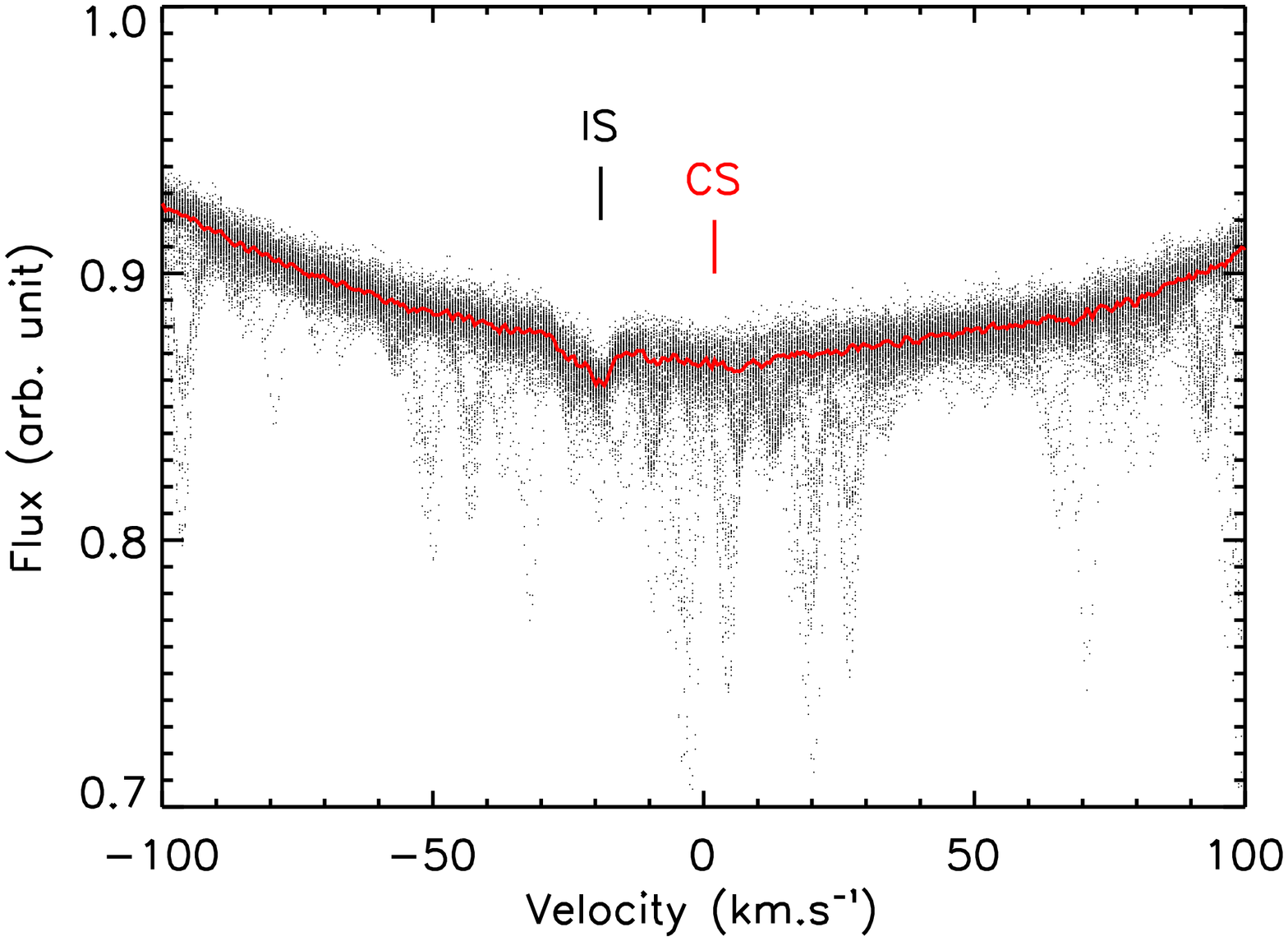}
\includegraphics[angle=90,width=0.48\columnwidth]{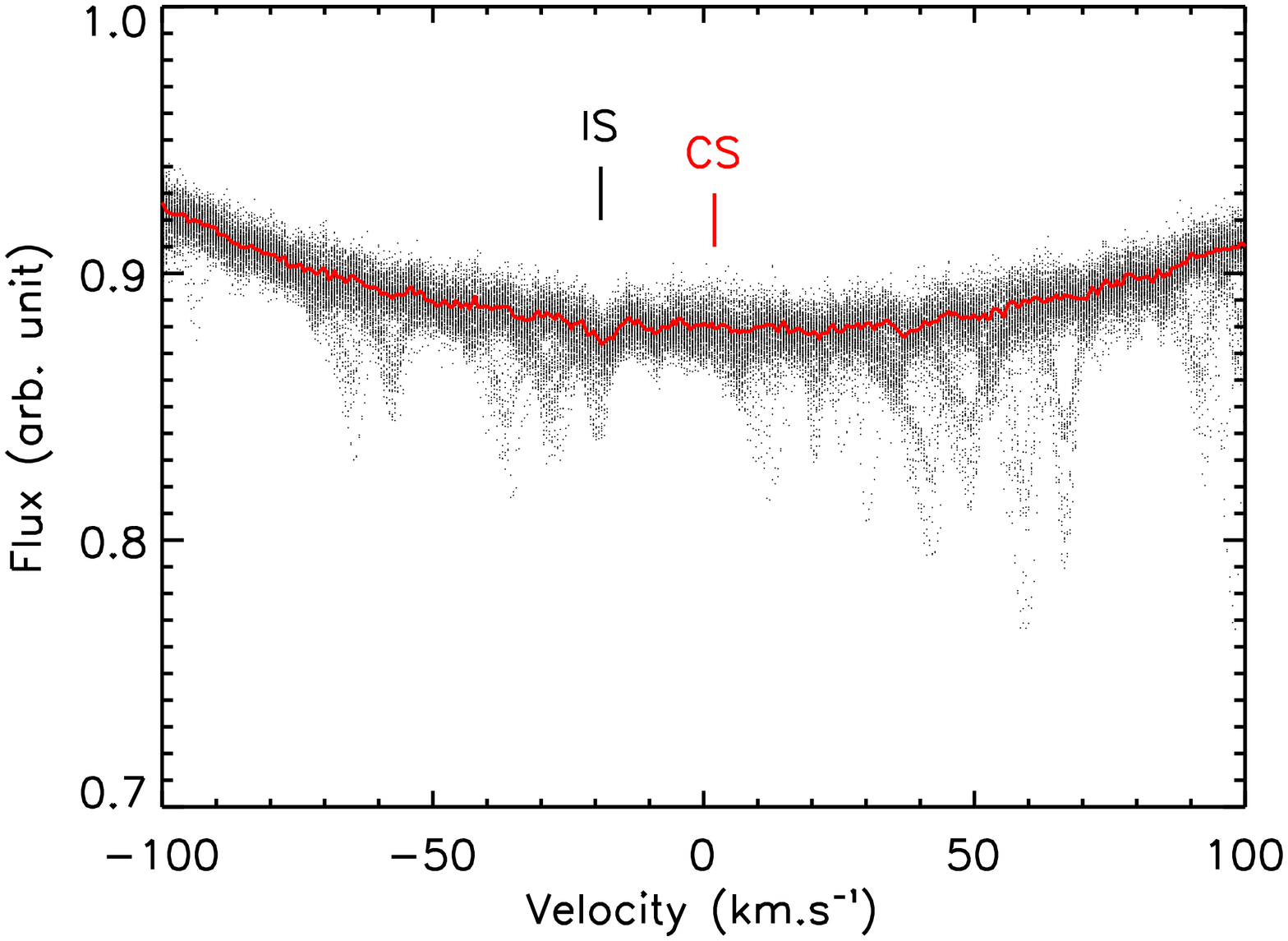}
}
\caption[Na I spectra]{
\label{fig:sodium}
Na I reference spectrum of HD172555 (red line) in the Na I D2 and D1 lines 
(left and right panels, respectively) using the same caption as that of Fig.~\ref{fig:ref_spec}. 
Velocities are expressed in the heliocentric rest frame. 
No significant Na\,I absorption feature appears at the star's radial velocity ($v$$\sim$2\,km\,s$^{-1}$), 
suggesting the CS origin of the absorption detected in the Ca\,II lines at this velocity. 
Conversely, the strongest absorption in Na\,I is seen at $v$$\sim$-19\,km\,s$^{-1}$ 
and is most likely of interstellar origin. The narrow absorption lines seen in most of the spectra 
(black dots) and not in the calculated reference spectrum are due to atmospheric water. 
}
\end{figure}
 
In the case of $\beta$\,Pic, this method has proved to be extremely robust for calculating 
the reference spectrum despite the presence of a large number of FEB absorptions in
most of the spectra (Kiefer et al.\ 2014, in preparation). 
The resulting reference spectrum for HD172555 is plotted in Fig.~\ref{fig:ref_spec}, 
where it is compared to the whole set of flux data points.

\section{Results}
\label{Results}

\subsection{The stable gaseous component}
\label{The stable gaseous component}

A first absorbing component in the Ca\,II appears at blueshifted velocities
($v\sim$-19\,km\,s$^{-1}$);
it is also clearly detected in the Na\,I D2 and D1 lines (Table~\ref{tab:EqW}).
A second feature is seen in the Ca\,II domain; this is the strongest feature
in the spectrum of HD172555. It is located
at the star radial velocity ($\sim$2\,km\,s$^{-1}$); no feature is detected
at the corresponding wavelengths in the Na\,I domain (Fig.~\ref{fig:sodium}).

The first feature at -19\,km\,s$^{-1}$ presents an
equivalent width ratio $EqW(\text{Na\,I}\,D_2)/EqW(\text{Ca\,II}\, K)$$\sim$1
which is typical of the local interstellar medium (Welsh et al.\ 2010).
In the line of sight of HD172555 (l=330.6\degr; b=-23.8\degr), an absorption by the G cloud
of the local interstellar medium is expected at -17\,km\,s$^{-1}$ (Lallement \& Bertin 1992;
Redfield \& Linsky 2008). We thus conclude that this stable component is most likely due to the G cloud in the interstellar medium.

On the contrary, the second feature is (i) seen at the stellar radial velocity,
(ii) the strongest absorption in the Ca\,II lines, and (iii) not seen in Na\,I.
This non-detection yields a 5-$\sigma$ upper limit for the equivalent widths
ratio of $EqW(\text{Na\,I}\,D_2)/EqW(\text{Ca\,II}\, K)$$<$0.04.
This upper limit is much lower than typical values
seen in the local interstellar medium where it is usually greater than 0.1 (see, e.g., Welsh et al.\ 2010).
Keeping in mind that the Na\,I/Ca\,II ratio
for the $\beta$\,Pic CS gas disk is also anomalously low compared
to IS standard, we conclude that the absorption detected in Ca\,II
at the star's radial velocity is most likely related to the CS gas of the HD172555 disk.

\subsection{Falling Evaporating Bodies}
\label{FEB}

\begin{figure}[htb]
\vbox{\centering
\includegraphics[angle=90,width=0.8\columnwidth]{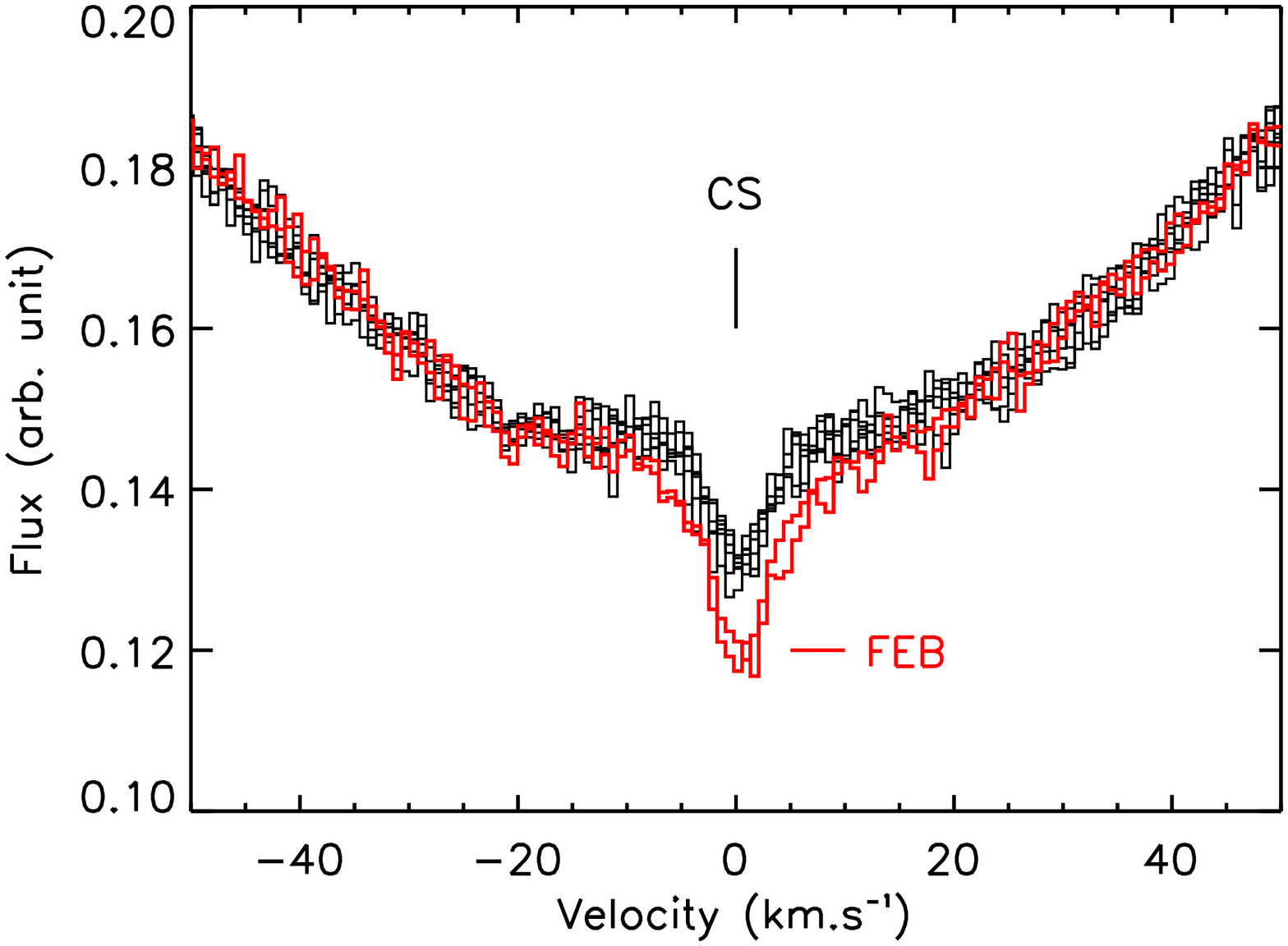}
\includegraphics[angle=90,width=0.8\columnwidth]{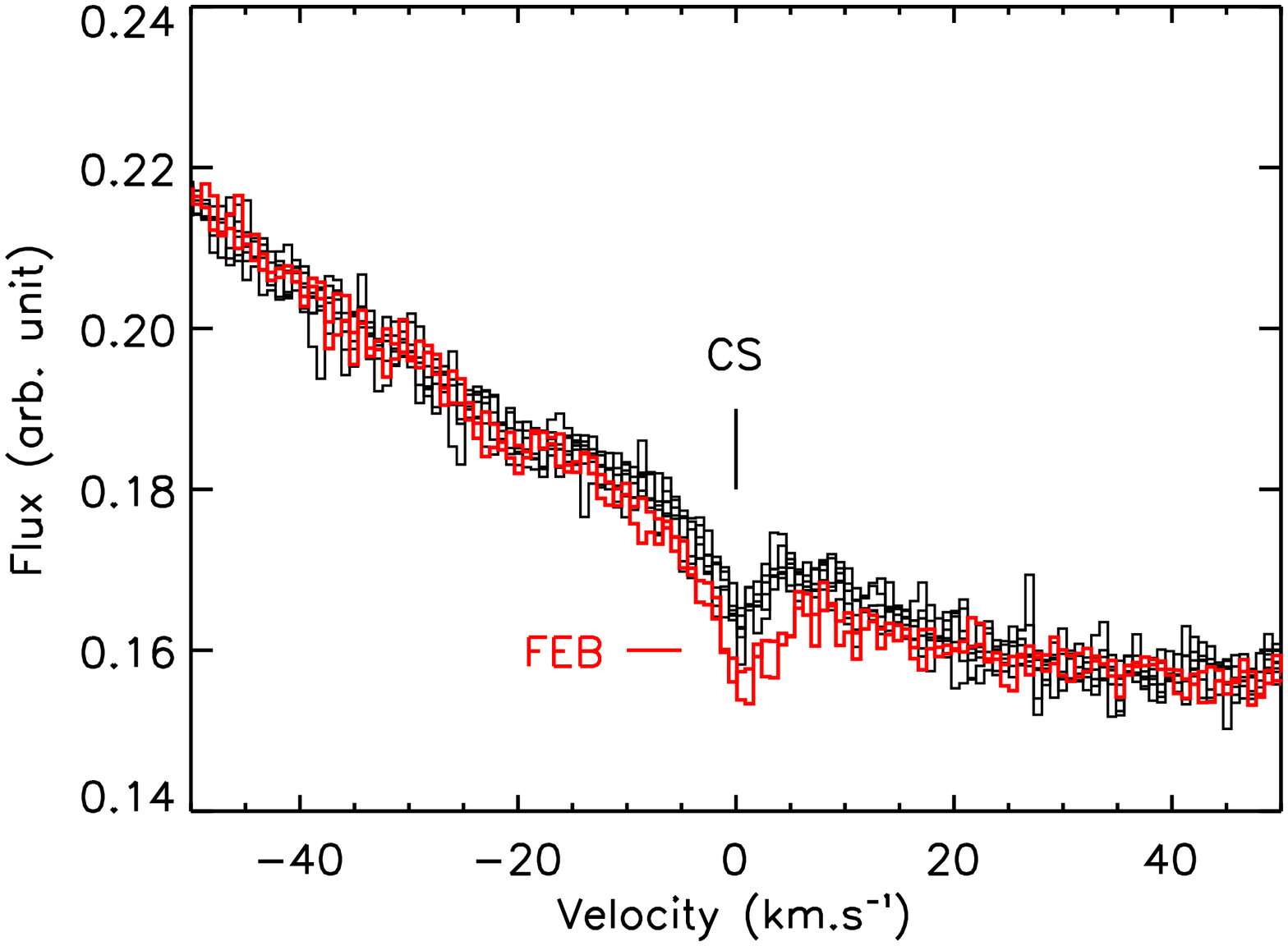}
}
\caption
[Some HD172555 spectra]
{
\label{fig:spec_comp}
Plot of HD172555 spectra of 2004 in the Ca\,II K line (top panel) and H line (bottom panel). 
Two typical spectra on 22 September 2004 separated by a 6 minute-time interval (red lines) clearly show the presence 
of an additional absorption at the star's radial velocity in comparison to spectra obtained 
100~days earlier (black lines). Velocities are given in the star's rest frame.
}
\end{figure}

A quick look at the \emph{HARPS} HD172555 spectra immediately reveals the presence of variations 
in the CS Ca\,II line seen at the star radial velocity and identified as due to CS gas 
(Sect.~\ref{The stable gaseous component}). These variations are seen at four different 
epochs over a total of 22~nights of observations (Table~\ref{tab:time_tab}). 
They always show up as additional absorptions over the 
main CS absorption, and their detection in both the Ca\,II-K and Ca\,II-H lines confirms that they
are true additional absorptions to the quiet spectrum (Fig.~\ref{fig:spec_comp}).

\begin{table}[tb]
\begin{tabular}{lccccc}
\hline
Year & \multicolumn{2}{c}{Date  } & Nb of & Comment & Minimum\\
     & (MJD) &(D/M/Y) & spectra &  &  duration \\
\hline 
2004 & 53147 & & 6 &  &  \\
     & 53156 & & 3 &  &  \\
     & 53270 & 22/09/04 & 6 & FEB & 3.5~h \\
\hline
2005 & 53603 & 21/08/05 & 2 & FEB & 3~min \\
\hline
2006 & 53875 & & 2 &  &  \\
     & 53880 & & 2 &  &  \\
     & 53881 & & 4 &  &  \\
     & 53989 & & 2 &  &  \\
\hline
2009 & 54942 & & 15 &  &  \\
     & 54943 & & 10 &  &  \\ 
     & 55032 & & 1  &  &  \\
     & 55035 & & 1  &  &  \\
     & 55148 & & 2  &  &  \\
\hline
2010 & 55384 & & 2  &  &  \\
     & 55385 & 08/07/10 & 2  & FEB & 3~min \\
\hline
2011 & 55723 & 11/06/11 & 31 & FEB & 1.9~h \\
     & 55724 & & 2  &  &  \\
     & 55725 & & 2  &  &  \\
     & 55726 & & 4  &  &  \\
     & 55763 & & 2  &  &  \\
     & 55764 & & 2  &  &  \\
     & 55765 & & 26 &  &  \\
\hline
\end{tabular}
\caption[Events timetable]{
\label{tab:time_tab} 
Harps observations of HD172555. Minimum duration is the time separating the event's first and last detections, i.e. the upper and lower bounds to the comet's transit ingress and egress times, respectively.
}
\vspace{-0.4cm}
\end{table}

These variations have time scales that are probably shorter than one day since, as can be seen in Table~\ref{tab:time_tab}, their occurrence is quite rare and at least one event, on 11 June 2011, is not observed the next day; 
a similar example also occurred on the 8 July 2010, but was not seen the night before. 
With such a low occurrence rate, detecting an event that lasted several days 
just before it disappears or appears would be very fortuitous. 

These variations thus share many common characteristics with those typically detected in $\beta$\,Pic spectra and identified as transits of exocomets (or FEBs). 
To quantify the characteristics of these FEBs features, we normalized the Ca II spectra by dividing the observed spectra by the reference spectrum. 
Then the absorption features are fitted by a toy-model of the absorbing gaseous cloud given by
\begin{equation}
\varepsilon(\lambda) = 1 - \alpha \left( 1 - e^{-\tau(\lambda)}\right),
\nonumber
\end{equation}
where $\alpha$ is the fraction of the stellar surface $\Sigma_\star$ covered by the cloud 
with an area $\Sigma_c$ ($\alpha$=$\Sigma_c$/$\Sigma_\star$$\le$1). 
The optical depth $\tau(\lambda)$ at the wavelength $\lambda$ is given by 
\begin{align}
\tau(\lambda) &= A e^{-\frac{(\lambda-\lambda_0)^2}{2\sigma^2}} \nonumber,
\end{align}
where the absorption depth $A$ depends on the absorbing medium density and thickness, 
and the line oscillator strength. Because the oscillator strength of the Ca\,II K line is twice that of the Ca\,II H line, the absorption depths in these lines follow the relationship $A_K=2 A_H$.

\begin{figure}[htb]
\hbox{
\includegraphics[angle=90,width=0.48\columnwidth]{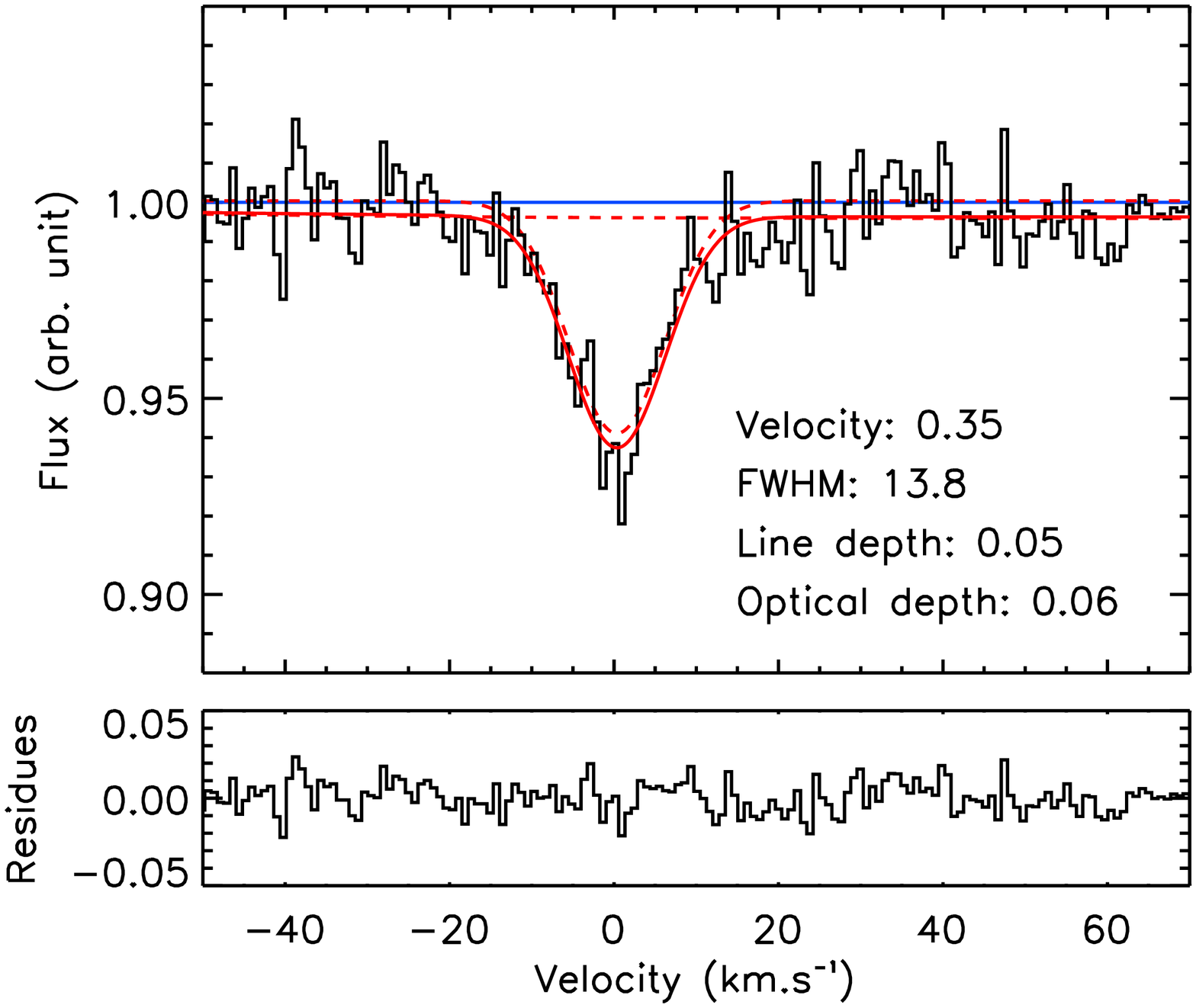}
\includegraphics[angle=90,width=0.48\columnwidth]{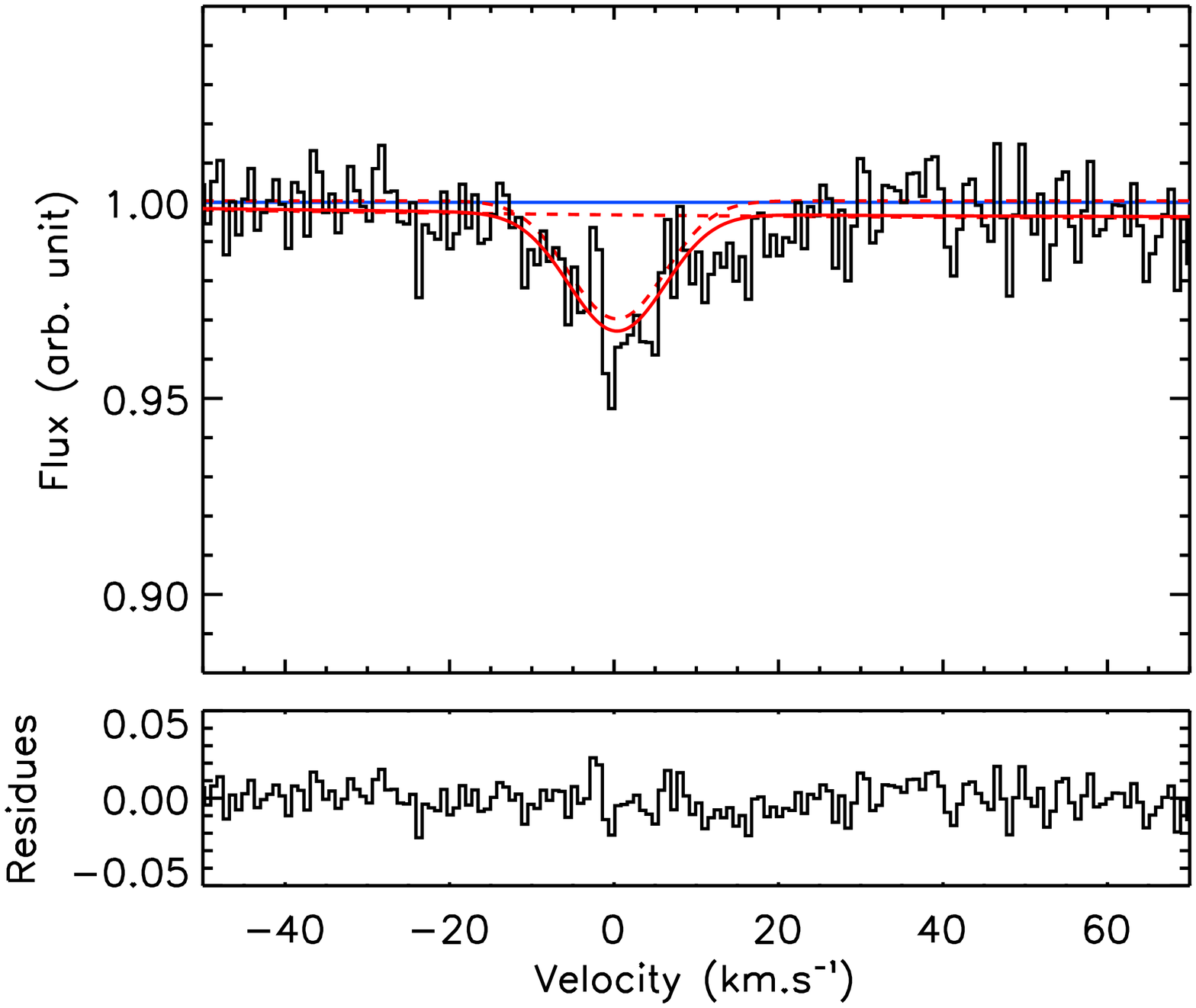}
}
\caption[Ca II spectrum, 22/09/04]{
\label{fig:abs_spec 53270}
Ca II normalized spectrum of the FEB detected on 22 September 2004 (MJD 53269.996). The Ca\,II-K and Ca\,II-H lines are plotted in the left and right panels, respectively. The fits to the data are plotted with the red lines. 
The bottom panels show the residuals of the fits.
}
\end{figure}

\begin{figure}[htb]
\hbox{
\includegraphics[angle=90,width=0.48\columnwidth]{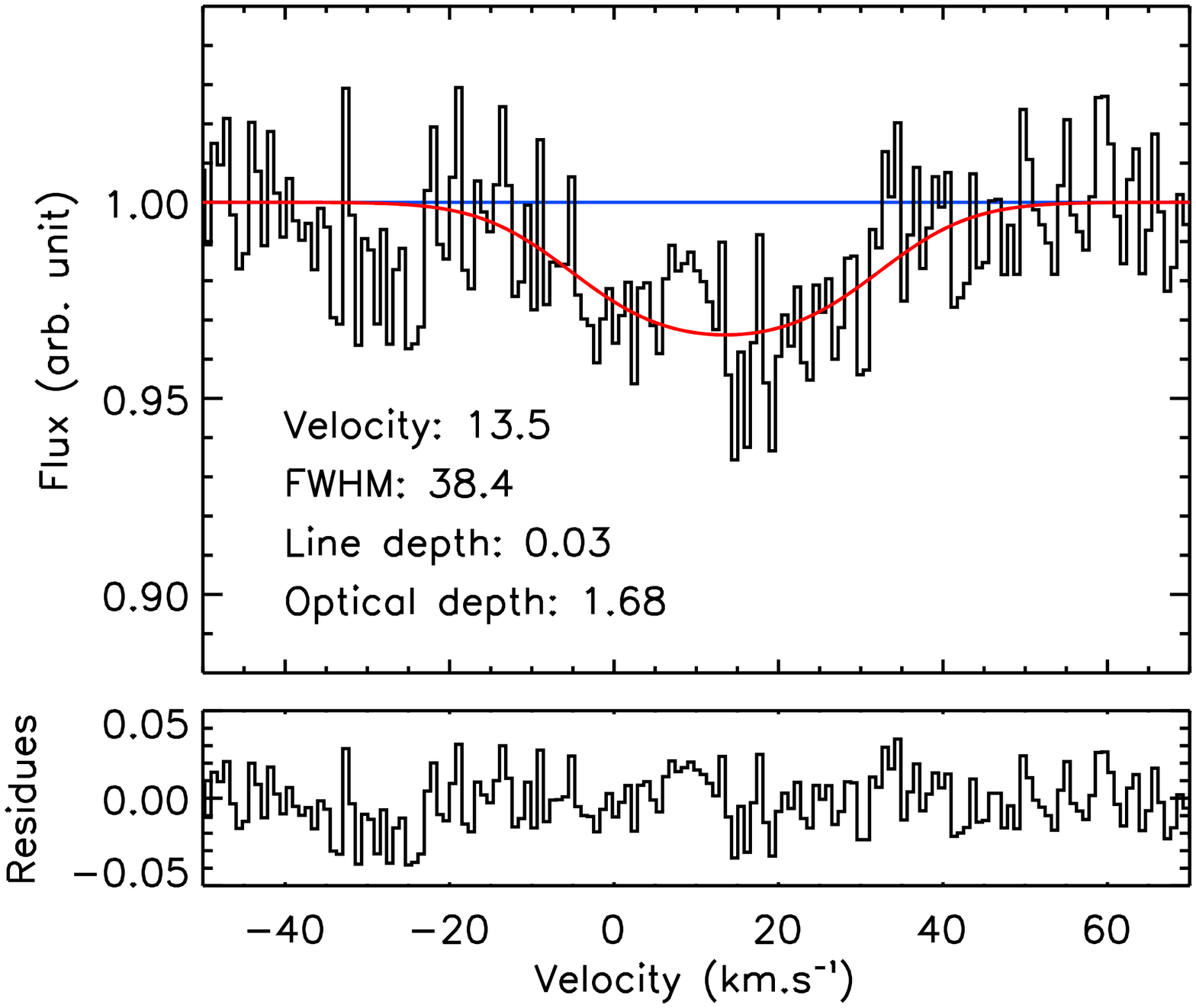}
\includegraphics[angle=90,width=0.48\columnwidth]{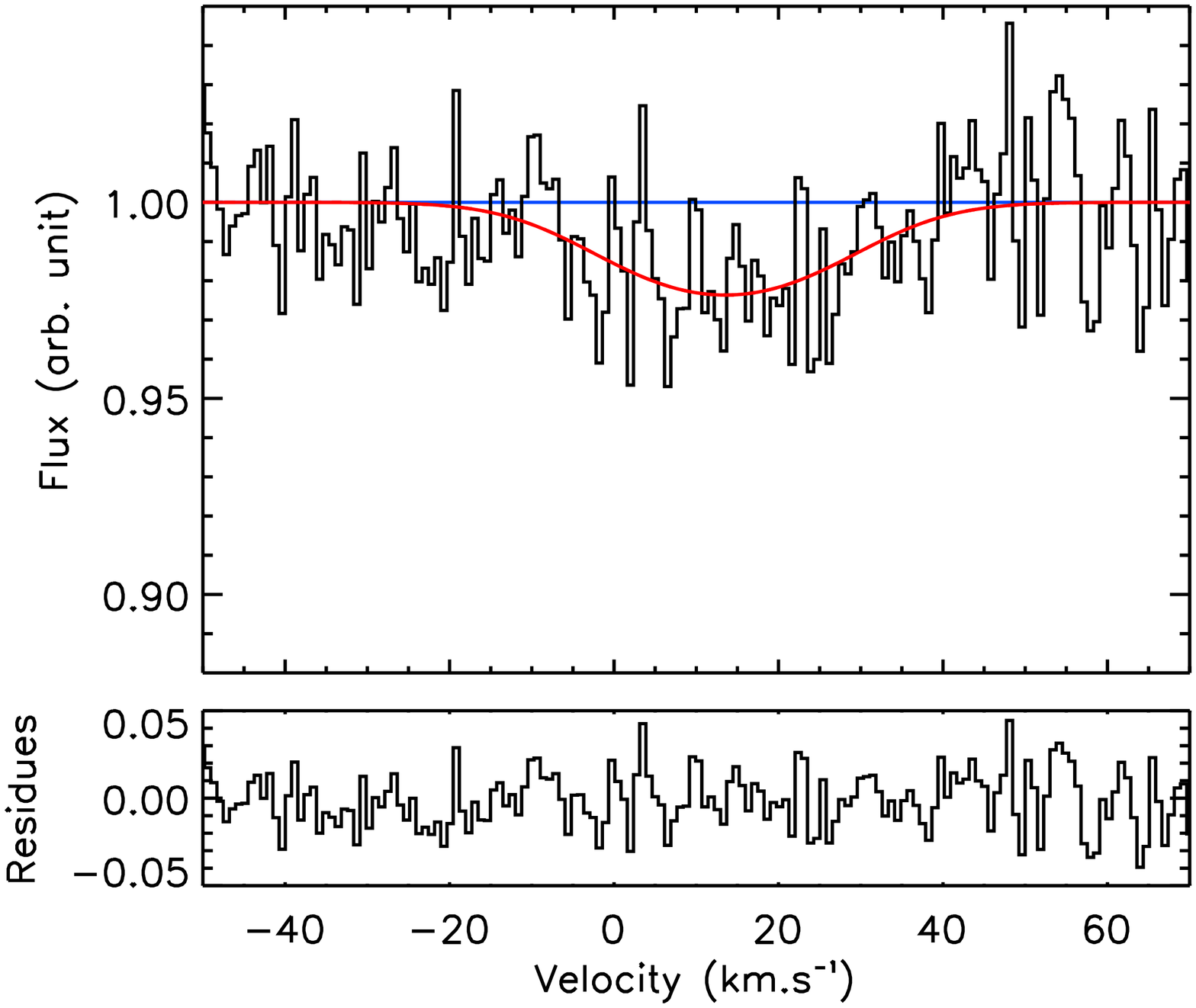}
}
\caption[Ca II spectrum, 21/08/05]{
\label{fig:abs_spec 53603}
Ca II normalized spectrum of the FEB detected on 21 August 2005 (MJD 53603.145), as in Fig.~\ref{fig:abs_spec 53270}.
}
\end{figure}

\begin{figure}[htb]
\hbox{
\includegraphics[angle=90,width=0.48\columnwidth]{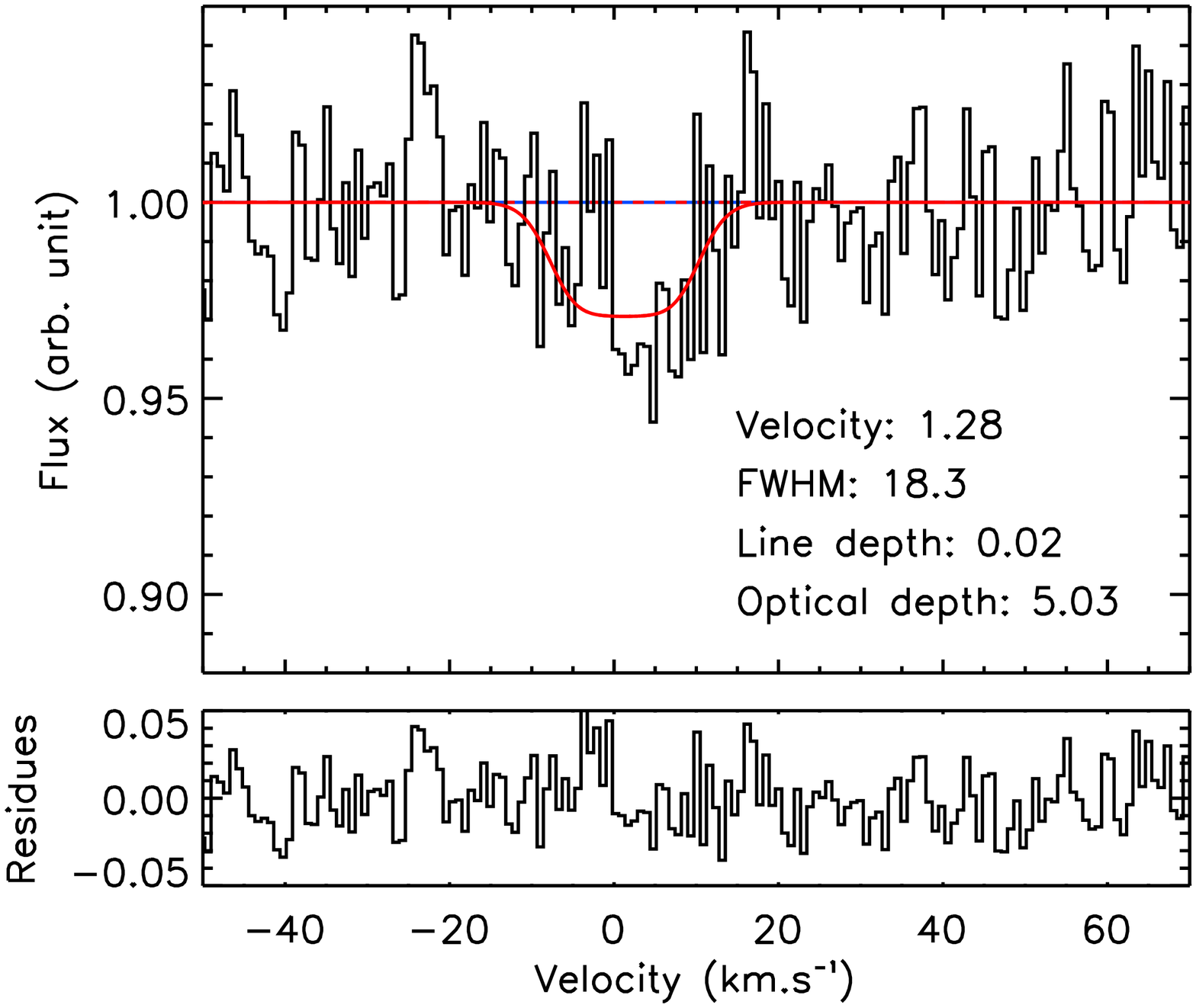}
\includegraphics[angle=90,width=0.48\columnwidth]{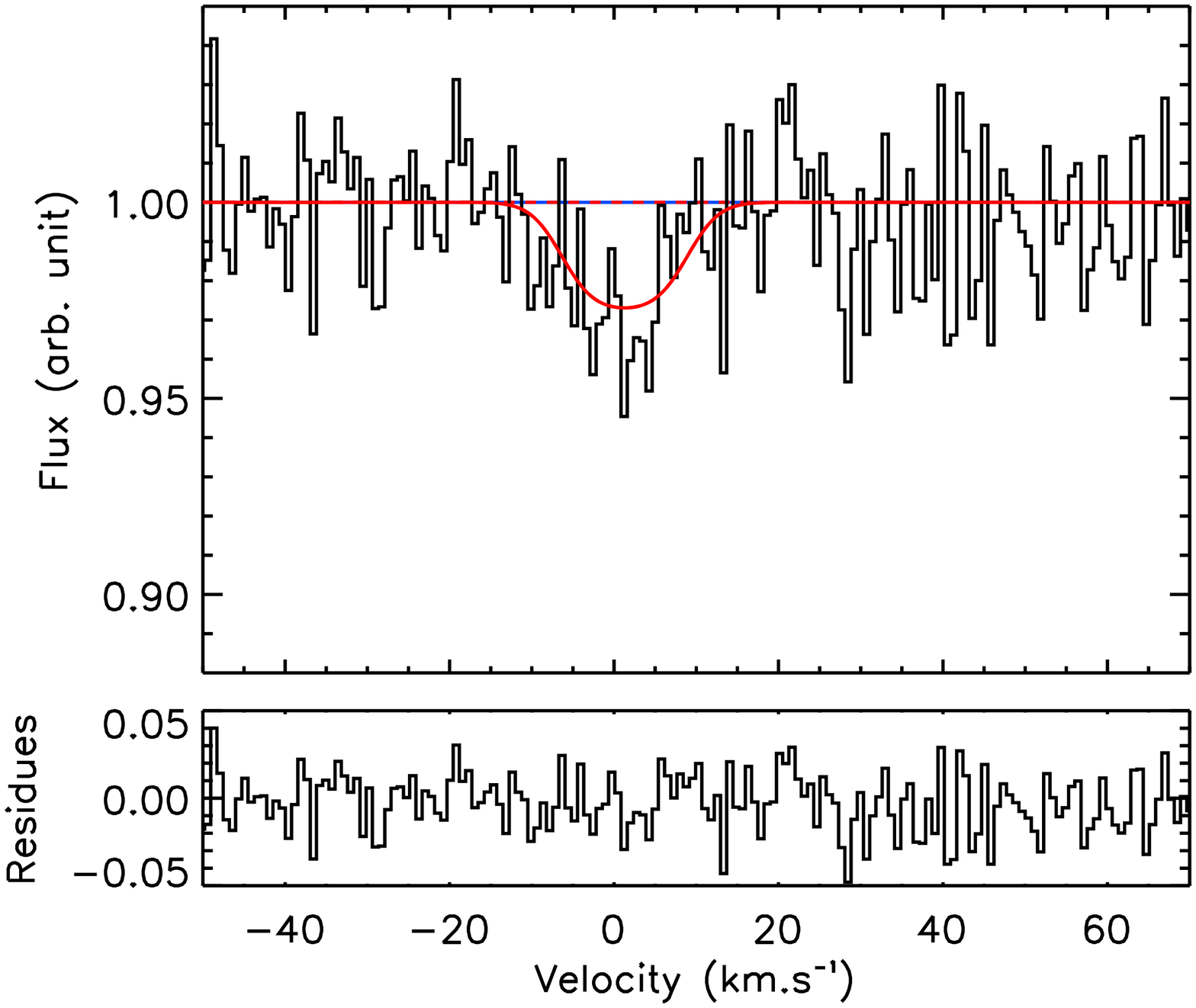}
}
\caption[Ca II spectrum, 08/07/10]{
\label{fig:abs_spec 55385}
Ca II normalized spectrum of the FEB detected on 08 July 2010 (MJD 55385.285), as in Fig.~\ref{fig:abs_spec 53270}. 
}
\end{figure}

\begin{figure}[htb]
\hbox{
\includegraphics[angle=90,width=0.48\columnwidth]{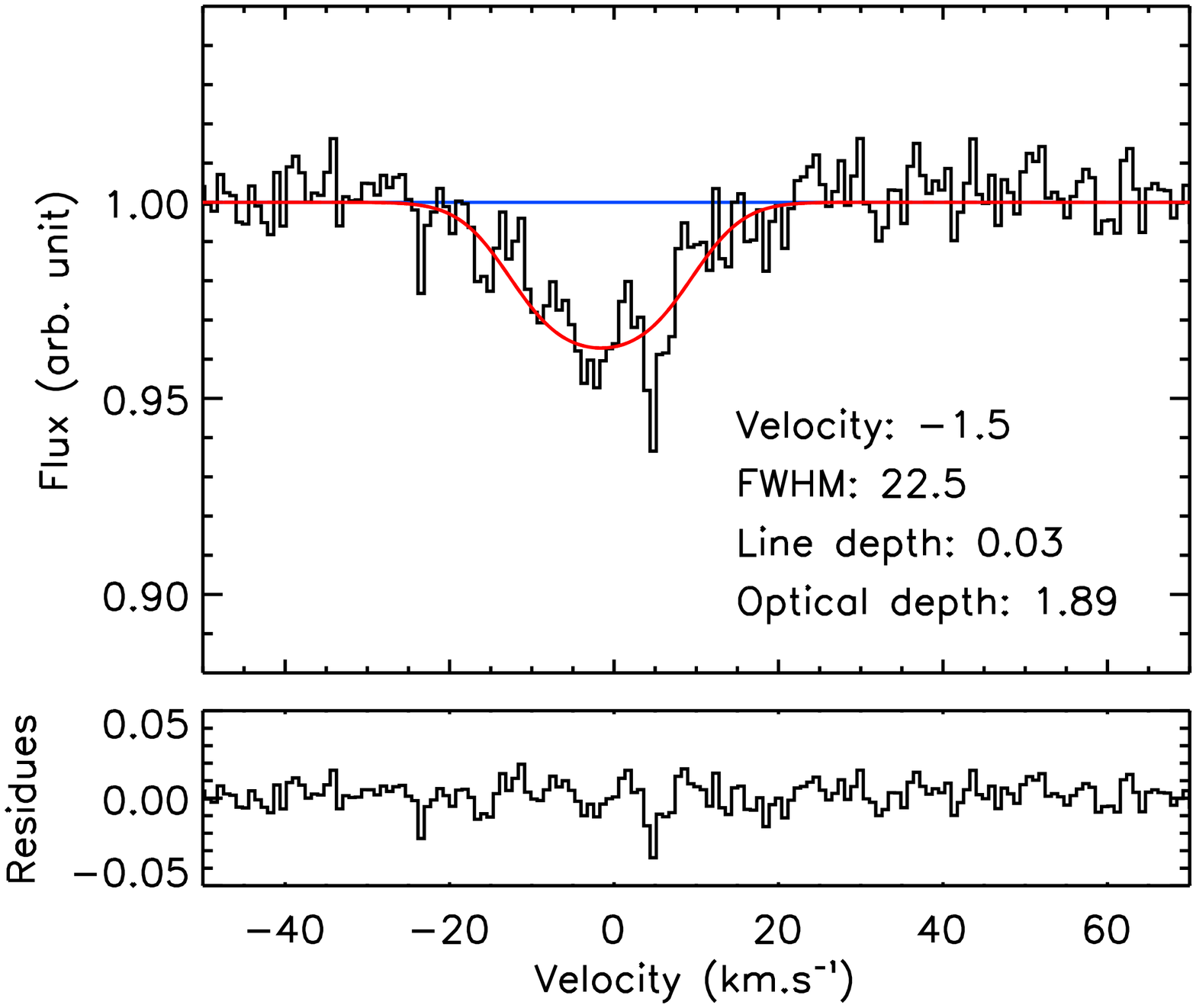}
\includegraphics[angle=90,width=0.48\columnwidth]{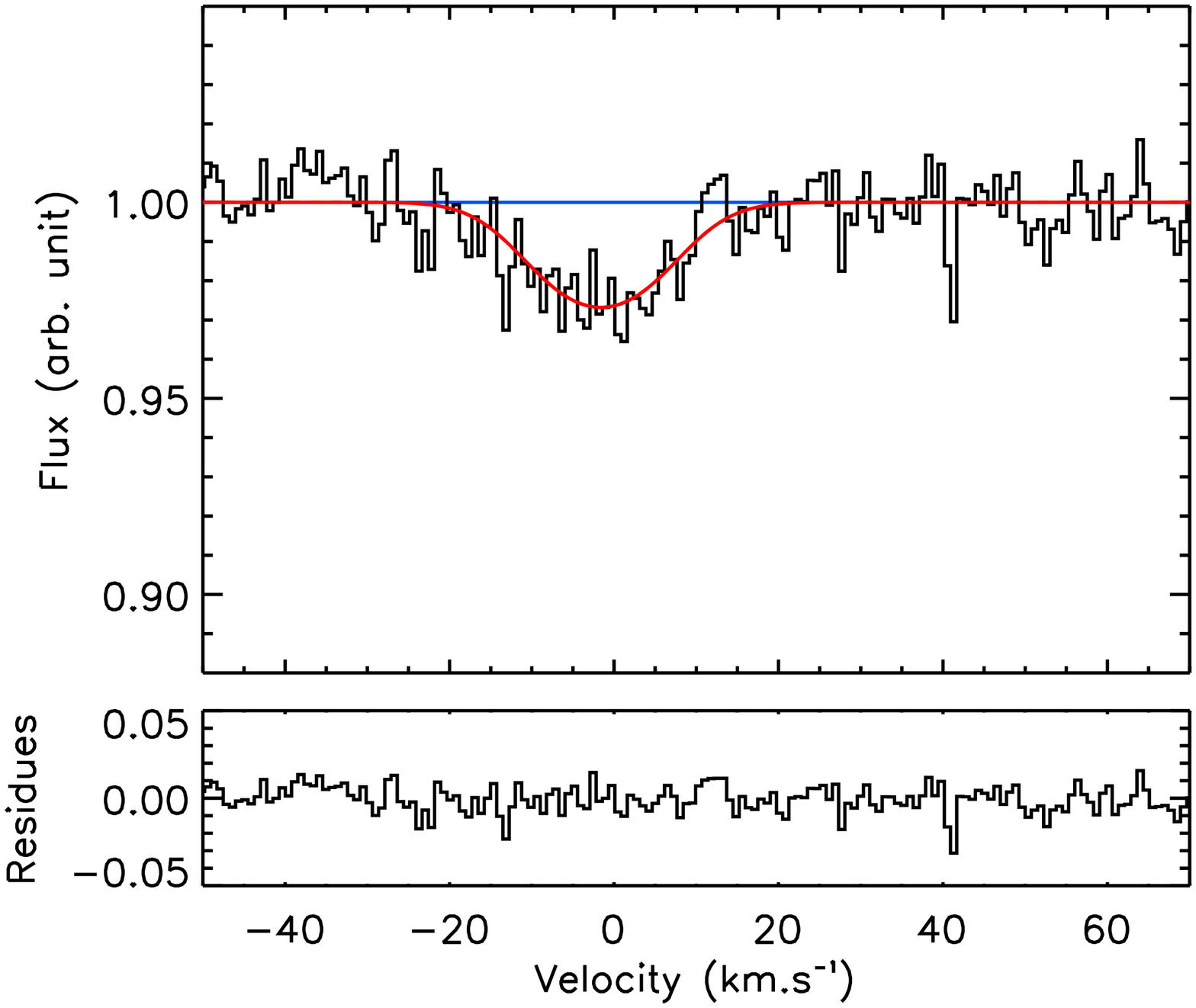}
}
\caption[Ca II spectrum, 11/06/11]{
\label{fig:abs_spec 55723}
Ca II normalized spectrum of the FEB detected on 11 June 2011 (MJD 55723.248), as in Fig.~\ref{fig:abs_spec 53270}.
}
\end{figure}

\begin{table*}[tb]
\begin{center}
\begin{tabular}{lcccccc}
\hline
\multicolumn{2}{c}{Date  }  & K line Depth & Velocity  & FWHM  & Surface Ratio & Optical depth \\
(MJD) & (D/M/Y)  & &  (km/s) & (km/s) & $\alpha$ & \\
\hline
53269.996 & 22/09/04 & $0.059 \pm 0.003$ & $0.35 \pm 0.37$ & $13.9 \pm 0.9$ &
$\ga 0.9$ &  $0.061 \pm 0.003$ \\
53270.134 & 22/09/04 & $0.072 \pm 0.006$ & $2.3 \pm 0.5$ & $19.5 \pm 1.2$ &
$\ga 0.84$ & $0.075 \pm 0.006$ \\
53603.145 & 21/08/05 &$0.034 \pm 0.004$ & $13.5 \pm 1.5$ & $38.4 \pm 9.7$ &
$0.04^{+0.04}_{-0.01}$ & $1.69 \pm 1.05$  \\
55385.285 & 08/07/10 & $0.029 \pm 0.006$ & $1.26 \pm 1.07$ & $18.4^{+7.9}_{-10.1}$
& $\ga 0.024$ & $\la 10.3$  \\
55723.248 & 11/06/11 & $0.037 \pm 0.002$ & $-1.6 \pm 0.5$ & $22.5 \pm 3.2$ &
$0.04\pm 0.01$ & $1.90 \pm 0.63$ \\
55723.280 & 11/06/11 &$0.032 \pm 0.002$ & $-2.44 \pm 0.51$ & $24.9 \pm 3.3$ &
$0.04 \pm 0.01$ & $1.48 \pm 0.51$ \\
55723.309 & 11/06/11 &$0.030 \pm 0.003$ & $-3.24 \pm 0.73$ & $24.2 \pm 4.7$ &
$0.04^{+0.02}_{-0.01}$ & $1.59 \pm 0.78$  \\
\hline
\end{tabular}
\caption[Fit parameters with 1-sigma error bars]{
\label{tab:param_tab}
Table of the fit parameters with 1-sigma error bars.
The upper and lower limits are given at the 1-sigma level. 
}
\end{center}
\vspace{-0.5cm}
\end{table*}

The normalized spectra of the four epochs with positive detection of FEBs are plotted
in Figs.~\ref{fig:abs_spec 53270}-\ref{fig:abs_spec 55723}. To increase 
the signal-to-noise ratio, we averaged the spectra taken over 30 minute-time intervals.  
The results of the fits are given in Table~\ref{tab:param_tab}. 

The FEBs are mostly seen at the star radial velocity, except for the FEB on 21 August 2005 with a 
redshifted velocity of $\sim$13\,km\,s$^{-1}$ (Fig.~\ref{fig:abs_spec 53603}). 
The most significant events on 22 September 2004 and 11 June 2011 have very different cloud characteristics, 
as often observed in $\beta$\,Pic (Lagrange et al.\ 1992). 
On 22 September 2004 the absorption in the Ca\,II-K line is about two times deeper than in the Ca\,II-H line
(Fig.~\ref{fig:abs_spec 53270}), 
in agreement with the ratio of the lines' oscillator strengths. This is easily explained if the
features are produced by large translucent clouds covering the whole surface of the star during the transit. 
On 11 June 2011 the absorption in the Ca\,II-H line is almost as deep as in the K line 
(Fig.~\ref{fig:abs_spec 55723}); 
this can be explained only if it is produced by an optically thick cloud, 
covering only a fraction of the star surface ($\alpha$$\sim$0.04$<$1). The ratio of the lines can be equal to one if they are produced by optically thick clouds (see discussion 
in Lecavelier des Etangs et al.\ 1997).

Although the two events on 21 August 2005 and 08 July 2010 are noisy, their detection remains statistically significant, 
thanks to the simultaneous presence of the absorption both in the 
Ca\,II-K and Ca\,II-H lines. 
For comparison, the very nice signal detected on 11 June 2011 was actually obtained by averaging 
around 10~spectra similar to the ones seen on 21 August 2005 and 08 July 2010.
Because of the noise, our line-finder procedure detected only one statistically significant FEB on 21 August 2005 (Fig.~\ref{fig:abs_spec 53603}); 
however, the plot suggests the possibility of two~absorbing components: one 
at about the star's radial velocity and another at~20~km/s. 
This would explain the particularly large FWHM 
found by fitting the line with a single component (Table~\ref{tab:param_tab}).

Table~\ref{tab:param_tab} also shows that the small thick clouds are found together with large FWHM. 
In the framework of the FEB scenario, this is interpreted by clouds transiting at very short distances 
to the star (a few stellar radii) for which large orbital velocity dispersion within the FEB coma yields 
large Doppler broadening of the lines. Meanwhile, high radiation pressure compressed the extension 
of the exocomets' coma into very small size (Beust et al.\ 1991, 1996). 
These similarities with the $\beta$\,Pic FEBs strengthen the conclusion 
that we are witnessing the transits of exocomets in the young circumstellar disk of HD172555, 
like in $\beta$\,Pic.

\section{Discussion}
\label{Discussion}

Considering that HD172555 and $\beta$\,Pic have similar ages and spectral types, and that both stars 
are surrounded by gaseous and dusty circumstellar disks, the detection of (i) an absorption 
with an anomalously low Na\,I/Ca\,II ratio at the star radial velocity, and (ii) four events 
of variable absorptions in the Ca\,II line, it is very likely that we are witnessing in HD172555 
a similar phenomenon to the one in $\beta$\,Pic explained by the transit 
of comet-like evaporating bodies.

Although the statistics is still poor for HD172555, we can already note a difference in Doppler velocities between the $\beta$\,Pic and HD172555 FEBs. While in $\beta$\,Pic most of the FEB signatures appear to be redshifted 
up to a few hundreds of km/s (see review in Vidal-Madjar et al.\ 1998), the FEBs of HD172555 show up 
nearly at the star's radial velocity within $\pm$10\,km\,s$^{-1}$. Importantly, this is not in contradiction with the exocomet (FEB) 
scenario, where the velocities of the sporadic absorptions are defined by the projection on the line of sight 
of the comet's orbital velocity. Thus, redshifted and blueshifted absorptions correspond 
to FEBs transiting before and after the periastron, respectively. 
In HD172555, the FEBs at low radial velocity must be transiting near their periastron (see Beust et al.\ 1991 
for the simulation of this configuration).

The detection of exocomets' transits raised the issue of the inclination 
of the planetary system relative to the line of sight. 
Beust \& Valiron (2007) have shown that most FEBs would remain spectroscopically undetected if the $\beta$ Pic disk was inclined by more that a few tens of degrees. 
This also applies to HD172555 if the dynamical origin of the FEBs is the same. 
Thus, the present detection of both the FEBs and the stable Ca\,II absorption 
argues that as $\beta$ Pic the HD172555 debris disk is probably close to edge-on. This remains consistent with the recent IR data constraining its inclination 
to be I$>$47$^\circ$ (Smith et al.\ 2012) and with the large $v\sin i$ value $\sim$116\,km\,s$^{-1}$ of this A7V star. 

The spectroscopic detection of transiting exocomets has already been made in a few cases other than
$\beta$\,Pic, for HR\,10 (Lagrange et al.\ 1990), and for HD\,21620, HD\,42111, HD\,110411, and 
HD\,145964 (Welsh \& Montgomery 2013). Here for the first time, we have presented the detection of 
a CS absorption feature associated with variable features detected simultaneously in the 
two Ca\,II lines, which ensure the veracity of their detection. 

Now further monitoring of HD172555 is needed to obtain additionnal data 
and to constrain characteristics of these exocomets orbiting within 
a young planetary system. 
In analogy with $\beta$\,Pic, 
observations of signatures of these exocomets in other spectral lines at different wavelengths can be anticipated 
with future observations.

\begin{acknowledgements}

This work has been supported by an award from the {\it Fondation Simone et Cino Del Duca}.
The authors also acknowledge the support of the French Agence Nationale de la Recherche (ANR), under programs ANR-12-BS05-0012 Exo-Atmos and ANR-2010 BLAN-0505-01 (EXOZODI).

\end{acknowledgements}

\end{document}